\documentclass[%
preprint, 
 amsmath,amssymb, 
 aps, 
]{revtex4-2} 

\usepackage{epsfig,dsfont,amssymb,amsmath,amsthm,amsfonts,amsbsy,mathrsfs} 
\usepackage{graphicx} 
\usepackage{color} 
\usepackage{bm}
\usepackage{multirow} 
\usepackage{natbib}

\newcommand{\pr}{{\cal P}}

\newcommand{\beq}{\begin{equation}}
\newcommand{\eeq}{\end{equation}}
\newcommand{\bea}{\begin{eqnarray}}
\newcommand{\eea}{\end{eqnarray}}

\begin{document}

\title{Confined run-and-tumble swimmers in one dimension}

\author{Luca Angelani}

\address{
ISC-CNR, Institute for Complex Systems, and Dipartimento di Fisica, 
Universit\`a {\it Sapienza}, Piazzale Aldo Moro 2, 00185 Rome, Italy}


\begin{abstract}
The persistent character of the motion of active particles gives rise to
accumulation at boundaries. 
I investigate the problem of run-and-tumble swimmers confined in a 1D box with hard walls,
reporting expressions for the particles probability distribution and wall  pressure. 
A crossover box length value is found below which the initial value of the pressure turns 
out to be higher than the asymptotic one, indicating a {\it bounce} effect of the 
active ``wave'' of swimmers.
The case of attracting and repelling boundaries are also investigated 
using two different tumble rates for particles in the bulk and at walls.
Escape problems are finally analyzed by considering partially permeable walls
through which particles can leave the box.
\end{abstract}

\maketitle

\section{Introduction}

Active matter has the property to accumulate at boundaries
when constrained in confining environments.
Spermatozoa \cite{Rot_1963} and {\it E. coli} cells \cite{Berg,Fry_1995,Ber_2008}
have been observed to preferentially swim near surfaces.
Many theoretical and numerical works have confirmed the tendency of active
particles to accumulate at walls,
ascribing it to hydrodynamic effects or to self-propulsion plus steric interaction mechanisms
\cite{Li_2009,Elg_2013,Elg_2015,Ewg_2015,Ezh_2015,Solon_2015,Mag_2015,Wen_2008,Cos_2012,Kai_2012,Yan_2014,Sch_2015,Mol_2014,Bech_2016}.
This property is at the basis of many unusual effects observed in active systems, such as
the ratchet effects \cite{Gal_2007,ratchet_2011,Rei_2016}, 
the ability to power micro-machines \cite{Ang_2009,Rdl_2010,Sok_2010,Ang_2010}
or the shape-shifting of soft-vesicles filled by active particles \cite{Pao_2016,Tia_2016}.
Strictly related is also the concept of pressure, which strongly depends on the 
interactions between the active particles and the confining boundaries
\cite{Solon_2015,Yan_2014,Sol_2015,Tak_2014,Fily_2014,Smal_2015,Nik_2016}.
Obtaining exact solutions for the problem of confined active particles is then of great importance
to better understand the phenomenon and as a reference for experimental and numerical results.
Up to now formulae have been obtained only in the steady state regime \cite{Elg_2013,Elg_2015,Solon_2015}.
Here we 
present a formulation of the problem which allow us to 
solve the dynamical equation  (telegrapher's equation) for the probability distribution 
function of swimmers in one space dimension.
More specifically we 
analyze the problem of run-and-tumble particles 
\cite{sch_1993,tail,tail2,mart,cat_2012,ang_epl_2013,det_2014}
confined in a finite 1D box with hard-wall boundaries.
In previous works 
the cases of reflecting and partially reflecting boundaries were analyzed 
\cite{EPJE_2014,Ang_2015},
without taking into account surface accumulation mechanisms.
Here we 
consider particles that perform a straight line motion at constant speed ({\it run}) interrupted by a
random reorientation of the direction of motion ({\it tumble}), and, when impinging on the hard wall,
they get stuck pushing on the boundary until a tumble event reverses the swimming direction.
By introducing suitable boundary conditions 
we investigate the solutions of the problem, by focusing on the probability distribution of 
swimmers and the pressure exerted on the walls.
Surface particles accumulation and an interesting {\it bounce} effect are observed:
for small box length values (with respect to the swimmers mean free path) the first collisions on the wall 
of particles which start their motion at the center of the box produces a pressure which turns out to be higher 
than the steady long-time pressure. 
We also analyze the case of non-trivial interactions between boundaries and swimmers, resulting in 
effective attractions or repulsions. We implement such effects by considering that 
the tumble rate of particles at wall could be different from the tumble rate in the bulk.
Moreover, by considering permeable walls which allow the particles to escape from the confined region, 
it is possible to study escape problems, focusing on the distribution of the escape time and its mean value.

\section{Run-and-tumble particles in 1D box}
  
\begin{figure}[t!]
\includegraphics[width=0.95\linewidth] {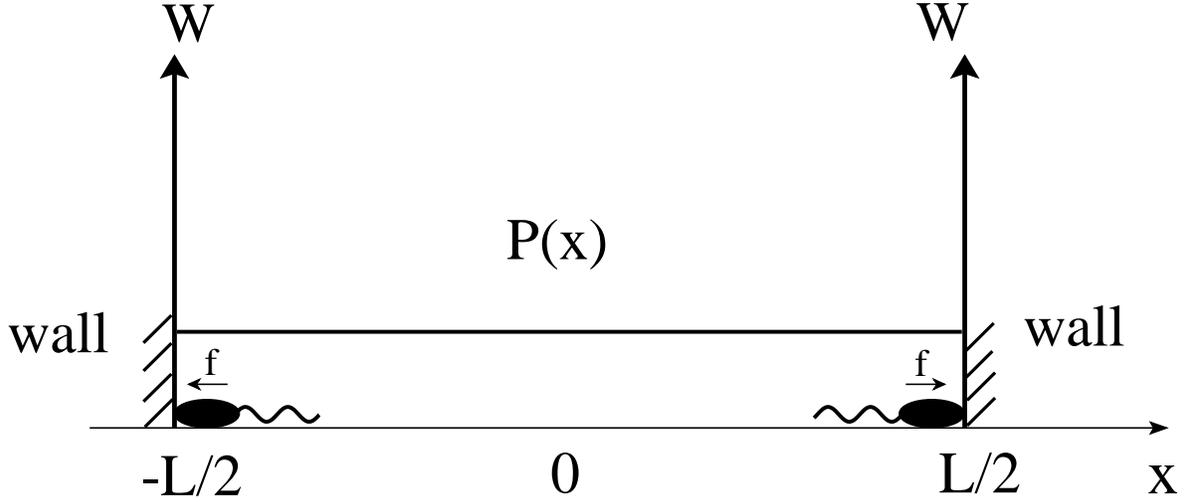}
\caption{\label{fig0}
Sketch of the problem of run-and-tumble particles confined in a 1D box with hard walls.
The showed density profile evidences the two peaks caused by accumulation of pushing swimmers at walls.
}
\end{figure}

Run-and-tumble particles in one dimension are described by the
following equations for the probability distribution functions of 
right-oriented and left-oriented particles, $\pr_{_R}(x,t)$ and $\pr_{_L}(x,t)$ 
\cite{sch_1993,tail,tail2,mart,cat_2012,EPJE_2014,Ang_2015}
\begin{eqnarray}
\label{eq_r}
\partial_t \pr_{_R} &=& - v \partial_x  \pr_{_R}
- \frac{\alpha}{2} \pr_{_R} + \frac{\alpha}{2} \pr_{_L} \\  
\label{eq_l}
\partial_t \pr_{_L} &=& v \partial_x \pr_{_L}
+ \frac{\alpha}{2} \pr_{_R} - \frac{\alpha}{2} \pr_{_L} 
\end{eqnarray}
where $\partial_t$ and $\partial_x$ are the time and space derivatives,
$v$ is the particle's speed and $\alpha$ the tumble rate,
i.e. the rate at which particles reorient their direction of motion.
The equations for the total PDF $\pr=\pr_{_R} + \pr_{_L}$ and current $J=v \pr_{_R} - v \pr_{_L}$
are
\begin{eqnarray}
\label{eq_p}
\partial_t \pr &=& - \partial_x  J \\  
\label{eq_j}
\partial_t J &=& - v^2 \partial_x \pr
- \alpha J 
\end{eqnarray}
We consider confining boundaries (hard walls) at $x=a$ and $x=b$
(without loss of generality we assume $a<0<b$).
Due to their persistent motion 
when particles reach the boundary they press against the wall 
until a tumble event reverses their direction of motion. 
This gives rise to particles accumulation at walls (see Fig. \ref{fig0} for a sketch of the problem).
We are considering here that the tumbling properties of swimmers are not affected by
the presence of the wall. Real bacteria could instead manifest a non-trivial 
dependence of tumble rates on the proximity to the boundary. 
Such a situation will be considered in Section 4.
We call $W_a(t)$ and  $W_b(t)$ the probabilities to find particles
stuck at  boundary points $a$ and $b$ at time $t$.
Normalization condition reads $\int_a^b dx\ \pr(x,t)+ W_a(t)+W_b(t)=1$.
The continuity equations for  $W_a$ and $W_b$ are given by
\begin{eqnarray}
\label{eq_wb}
\partial_t W_b(t) &=& J(b,t) \\  
\label{eq_wa}
\partial_t W_a(t) &=& -J(a,t)
\end{eqnarray}
We implement boundary conditions in the presence of hard walls as follow
\begin{eqnarray}
\label{eq_br}
v  \pr_{_L} (b,t) &=& \frac{\alpha}{2} W_b (t) \\  
\label{eq_bl}
v  \pr_{_R} (a,t) &=& \frac{\alpha}{2} W_a (t) 
\end{eqnarray}
obtained imposing that left (right) flux of particles at right (left) boundary point
is generated by the fraction of  particles stuck at boundaries that invert 
their direction of motion.

\noindent 
We assume that at initial time $t=0$ particles are concentrated at the origin $x=0$,
and are equally distributed between right and left oriented, i.e.
initial conditions are $\pr(x,0)=\delta(x)$ and $J(x,0)=0$
(and $W_a(0)=W_b(0)=0$).
Working in the Laplace domain, the resulting telegraph equation for 
${\tilde \pr}(x,s) = \int_0^\infty dt e^{-ts} \pr(x,t)$
-- derived from Eq.s (\ref{eq_p},\ref{eq_j}) -- is
\begin{equation}
v^2 \frac{\partial^2 {\tilde \pr}}{\partial x^2} - s (s+\alpha) {\tilde \pr} = -(s+\alpha) \delta(x)
\label{eq_lap}
\end{equation}
Boundary conditions at $x=a,b$ for ${\tilde \pr}$ are obtained from Eq.s (\ref{eq_br},\ref{eq_bl}) 
by using the relations among $\pr$, $J$ and $W$ in Eq.s (\ref{eq_j}) and (\ref{eq_wb},\ref{eq_wa}) in the Laplace domain:
\begin{eqnarray}
\label{eq_bpr}
v  \partial_x {\tilde \pr}|_b &=& - s {\tilde \pr}|_b \\  
\label{eq_bpl}
v  \partial_x {\tilde \pr}|_a &=&  s {\tilde \pr}|_a  
\end{eqnarray}
In the following we consider a symmetric situation, $b=-a=L/2$,
where $L$ is the box length.
The solution of Eq. (\ref{eq_lap}) is 
${\tilde \pr}=A_1 e^{c|x|} +A_2 e^{-c|x|}$, where 
\begin{equation}
v^2c^2(s) = s(s+\alpha)
\label{c}
\end{equation}
and 
$A_1=(c/2s)(vc-s)e^{-cL/2}/B$,
$A_2=(c/2s)(vc+s)e^{cL/2}/B$,
with $B= (s+vc) e^{cL/2} - (vc - s)e^{-cL/2}$.
\begin{figure}[t!]
\includegraphics[width=0.95\linewidth] {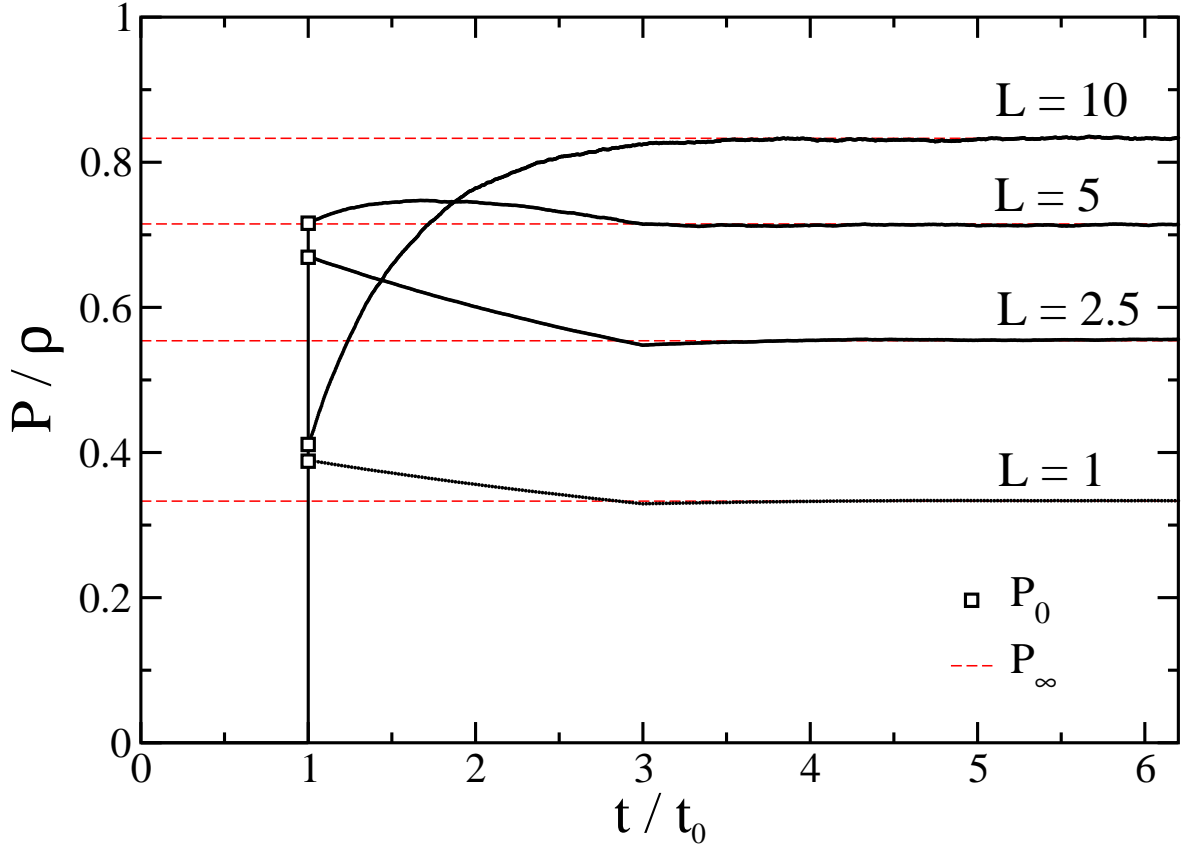}
\caption{\label{fig1}
Pressure $P$ (over particle density $\rho$) on the wall as a function of time for different box lengths. 
Times are rescaled to $t_0=L/2v$, which is the minimum time needed for a particle starting at the origin to reach the boundary.
Square symbols indicate the pressure $P_0$ at time $t_0$ and dashed-red lines are the asymptotic pressure values $P_\infty$.
Data obtained from simulations. 
Internal units are used: $v=1$, $\alpha=1$, $\mu=1$.
}
\end{figure}
We finally obtain the fraction of particles stuck at one boundary
(in the Laplace domain)
\begin{equation}
{\tilde W}(s) = \frac{1}{(vc+s) e^{cL/2} - (vc - s)e^{-cL/2}} 
= \frac{1}{2} \ \frac{1}{s\cosh{(cb)} + vc \sinh{(cb)}} 
\label{eq_q}
\end{equation}
For the probability distribution of particles  inside the interval we have
\begin{eqnarray}
{\tilde \pr}(x,s) &=& \frac{{\tilde W}(s)}{2v} \ \left[ (vc+s+\alpha) e^{c(L/2-|x|)} \right. \nonumber\\
&-& \left. (vc-s-\alpha) e^{-c(L/2-|x|)} \right]
\label{eq_px}
\end{eqnarray}
In the limit of $v,\alpha \to \infty$, keeping constant $v^2/\alpha$, the run-and-tumble particles become
Brownian (diffusive limit). 
In this case the fraction of particles at boundaries goes to 
zero, due to the fact that they are no more animated by persistent motion
and can not be found stuck at boundaries.
 Another interesting limit is obtained for small $\alpha$, approaching the non-tumbling case
(wave limit). 
For $\alpha=0$ one has ${\tilde W}(s)=e^{-Ls/2v}/2s$, that is, in the time-domain,
$W(t)=(1/2) \theta(t-t_0)$, with $\theta$ the
Heaviside step function:  particles starting at the origin reach the boundaries 
after a time $t_0=L/2v$ and stay there forever.

\noindent We note that the quantity $W(t)$ is different from zero only for $t>t_0=L/2v$,
due to the finite particles velocity.
We can then set  $W(t) = \theta(t-t_0) Q(t-t_0)$,
or, in the Laplace domain, ${\tilde W}(s)=e^{-Ls/2v}\ {\tilde Q}(s)$.
We can give an explicit expression for  the value of $W$ at $t_0\!=\!L/2v$, 
$W_0=W(t_0)=Q(0)$. 
By using the property 
$Q(0) = \lim_{s\to \infty} \ s {\tilde Q}(s)$,
one has
\begin{equation}
W_0 =  \frac12 \ \exp\left(-\frac{\alpha L}{4v}\right) 
\label{W0}
\end{equation}
where we have considered that, for large $s$, 
$vc \simeq s(1+\alpha/2s)$.

\noindent In the stationary regime we can  give an exact expression of $W$,
using the property $W_{\infty} = \lim_{t\to \infty} W(t)= \lim_{s\to 0} \ s {\tilde W}(s)$, which leads to
\begin{equation}
W_{\infty} = \frac{1}{2} \ \left( 1+\frac{\alpha L}{2v}  \right)^{-1}
\label{Wst}
\end{equation}
Inside the box particles are uniformly distributed with $\pr_{\infty} = \alpha W_{\infty} / v$.

\begin{figure}[t!]
\includegraphics[width=0.95\linewidth] {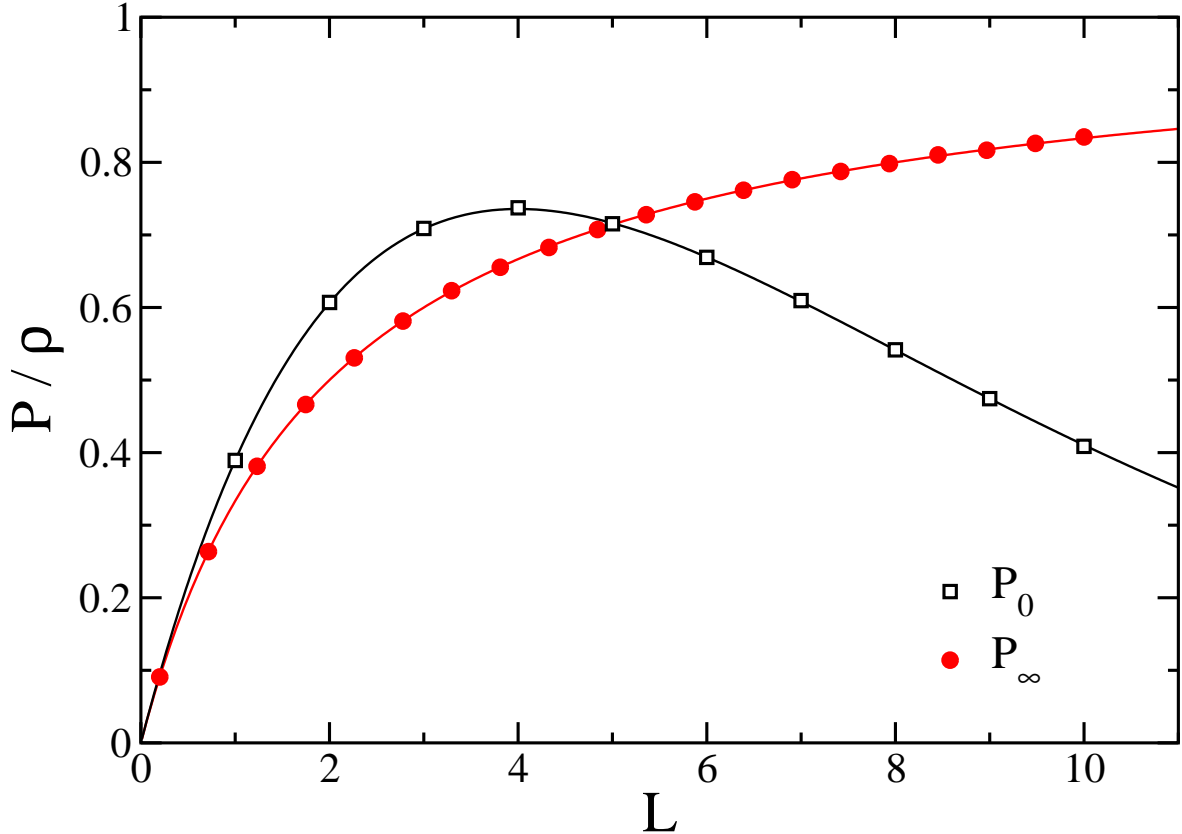}
\caption{\label{fig2}
Initial pressure $P_0$ and stationary pressure $P_\infty$ over particle density $\rho$ as a function of box length $L$.
Symbols are from simulations and lines from theory, Eq.s (\ref{P0}) and (\ref{Pinf}).
Internal units are used: $v=1$, $\alpha=1$, $\mu=1$.
}
\end{figure}

\section{Pressure}
The pressure exerted on the boundary by $N$ non interacting run-and-tumble particles confined in a 1D box of length $L$ ($\rho=N/L$)
can be easy obtained as
$P(t)=N W(t)\ f$, where $f$ is the force that each  particle exerts on the wall.
By considering hard walls, the force $f$  is exactly the force that the wall exerts on the particle and
it is the force required to hold the swimmer fixed (stall force).
In the overdamped regime the particle dynamics is described by 
$\dot{x} = v e + \mu f$, where $e=\pm 1$ is the orientation of the swimmer, $\mu$ the mobility and
$v$ the free swim velocity.
Particles stuck at boundary points ($\dot{x}=0$) exert then a force on the wall of strength $f=v/\mu$ and
the wall pressure can be obtained as $P(t)=N W(t)\   v/\mu$ \cite{Tak_2014,Fily_2014}.
We can give an analytic expression of the pressure in the Laplace domain by using the result of Eq. (\ref{eq_q})
\begin{equation}
{\tilde P}(s) = \frac{Nv}{\mu} \ \frac{1}{(vc+s) e^{cL/2} - (vc - s)e^{-cL/2}}
\label{P_s}
\end{equation}
We remind that this expression is valid for the case of particles starting their motion at the center of the box.
The wall pressure $P_0$ at time $t_0=L/2v$ and the asymptotic pressure at $t\to \infty$ $P_{\infty}$ 
can be  obtained from Eq.s (\ref{W0}) and (\ref{Wst}).
We have
\begin{equation}
P_0 =  \rho \frac{vL}{2\mu} \ \exp\left(-\frac{\alpha L}{4v}\right) 
\label{P0}
\end{equation}
and \cite{Mag_2015}
\begin{equation}
P_\infty = \rho \frac{v^2}{\mu \alpha} \ \left( 1+\frac{2v}{\alpha L}  \right)^{-1} 
\label{Pinf}
\end{equation}
The latter expression can be cast in the following form
\begin{equation}
P_\infty  \left( 1+\frac{2D}{Lv} \right)       = \rho k_B T
\label{eos}
\end{equation}
where $k_B T = D/\mu$ and $D=v^2/\alpha$.
The above expression can be considered as the ``equation of state'' for an ideal gas
of run-and-tumble particles in one space dimension.
It is worth noting that in spatial dimension higher than one it is not possible to define
an equation of state for generic active fluids, due to the existence of possible 
orientation-dependent interactions \cite{Solon_2015}.
In the Brownian limit, $v,\alpha \to \infty$ at constant $D$,
one recovers the equation of state of an ideal molecular gas,
$P = \rho k_B T$. 
The same result is obtained in the limit $L/\ell \gg 1$, i.e. 
for large box length with respect to the mean free path $\ell=v/\alpha$.

In Fig. \ref{fig1} we report the pressure as a function of time. 
The initial and asymptotic pressure values obtained from analytic expressions 
(\ref{P0}) and (\ref{Pinf})
are reported, respectively, as square symbols and dashed lines. Full lines in Fig. \ref{fig1}  
refer to numerical simulations,
performed as follow: we consider run-and-tumble particles
obeying the equation of motion $\dot{x}=ve$ 
where $e=\pm 1$ denotes the right/left swimming direction. 
Tumbling is implemented by randomly extracting a new direction $e$ with 
rate $\alpha$,
resulting in a Poissonian distribution of tumble times (times between tumble events).
When a particle reaches the boundary at $x=\pm L/2$, it gets stuck at that point until it 
changes its direction of motion.
The theoretical curves of pressure $P$ as a function of time, calculated by numerically 
inverting the Laplace tranformed distribution $\tilde{P}$ (\ref{P_s}), 
perfectly superimpose to simulation data.
As one can see from the figure, for box length values comparable to the particles mean free path $\ell$ 
($\ell=v/\alpha=1$ in internal units, $v=1$, $\alpha=1$, $\mu=1$),
the initial pressure $P_0$ is higher than the asymptotic one $P_\infty$,
while it becomes smaller at larger $L$.
We analyze the size dependence of the initial and asymptotic pressure in Fig. \ref{fig2}.
Interestingly, one observes a non-monotonic behavior of the initial pressure $P_0$ and a 
crossover at $L^*$
from a region where $P_0/P_\infty>1$, for $L<L^*$,  and a region where $P_0/P_\infty<1$, for $L>L^*$.
By equating the two expressions in Eq.s (\ref{P0}) and (\ref{Pinf}) we get
\begin{equation}
\exp\left( -\frac{\alpha L^*}{4v}\right) = \left( 1+\frac{\alpha L^*}{2v}  \right)^{-1} 
\end{equation}
giving rise to $L^* \simeq 5.03$ 
(in unit of $\ell=v/\alpha$).
We find then an excess  (up to $20\%$) of wall pressure at early time 
with respect to the asymptotic value,
when physical parameters are such that $L/\ell \leq 5.03$. 
This parameters region can be explored by varying the box length $L$ or the
swimmers features, such as the swim speed or the tumble rate, i.e. by considering different 
kinds of swimmers or light-powered bacteria \cite{rodops}.
It would be interesting to see if this is a peculiar finding of the present idealized and simplified
1D model of confined swimmers  or it is a general enough result that could be observed in real experiments.\\
We note also that real swimmers suspensions can manifest a large variability in some of 
the swimming parameters, such as run speeds or tumbling rates. 
In this case an exact expression for the pressure is obtained from previous results 
by averaging over speed or tumble rate distributions.
For example, by indicating with $f_v(v)$ the swimmers speed distribution, one has for the wall pressure
of an heterogeneous ensemble of non-interacting swimmers
$P = (N/\mu) \int dv f_v(v) v W(v)$, where $W(v)$ is the quantity previously obtained 
in the case of constant speed $v$
(a similar formula applies in the case of tumble rate distribution).

\section{Attractive and repulsive boundaries}

Boundaries can interact in a non-trivial way with structured swimmers. For example, it
has been numerically shown that flagellated bacteria can be both attracted or repelled by solid
walls depending on the bacterium shape \cite{Shum_2015}. {\it Escherichia coli} have been observed to
reduce their tumble rate by 50\% close to surfaces, preventing escape of bacteria from 
the boundaries \cite{Mol_2014}. 
In order to include in an effective way such bacteria-surface interactions in our simplified
1D model, we consider that the tumble rate of the swimmers at walls $\alpha_{_W}$ can be 
different from the bulk tumble rate $\alpha$. In such a way the boundary has the effect to 
modify, enhancing or suppressing, the ability of bacteria to change their swimming direction. 
The analysis is formally identical to that reported in the previous sections, but the boundary conditions
Eq.s (\ref{eq_br},\ref{eq_bl}) now read:
\begin{eqnarray}
\label{brw}
v  \pr_{_L} (b,t) &=& \frac{\alpha_{_W}}{2} W_b (t) \\  
\label{blw}
v  \pr_{_R} (a,t) &=& \frac{\alpha_{_W}}{2} W_a (t) 
\end{eqnarray}
or, in term of $\pr$ in the Laplace domain:
\begin{eqnarray}
\label{bprw}
v  \partial_x {\tilde \pr}|_b &=& - k(s) {\tilde \pr}|_b \\  
\label{bplw}
v  \partial_x {\tilde \pr}|_a &=&  k(s) {\tilde \pr}|_a  
\end{eqnarray}
where we have introduced the quantity
\begin{equation}
k(s)= \frac{s (s+\alpha)}{s+\alpha_{_W}}
\end{equation}
For $\alpha_{_W}=\alpha$ we have $k(s)=s$ and the problem reduces to the previously considered one.
By solving Eq. (\ref{eq_lap}) with the new boundary conditions, we finally
obtain the fraction of particles stuck at wall 
\begin{equation}
{\tilde W}(s) = \frac{1}{2 s} \  \frac{k(s)}{k(s) \cosh{(cL/2)} + vc \sinh{(cL/2)}}
\label{ww}
\end{equation}
Explicit expressions for the initial and stationary values are given by
\begin{equation}
W_0 =  \frac12 \ \exp\left(-\frac{\alpha L}{4v}\right) 
\label{W0w}
\end{equation}
and
\begin{equation}
W_{\infty} = \frac{1}{2} \ \left( 1+\frac{\alpha_{_W} L}{2v}  \right)^{-1}
\label{Wstw}
\end{equation}
The initial value $W_0$ is independent of $\alpha_{_W}$, as expected due to the fact that 
it is determined by 
first swimmers which reach the wall, that is the fraction of swimmers 
which travel from the origin to the wall in a time $t=L/2v$ without tumbling, 
and thus it does not depend on the wall properties.
For attracting walls,  $\alpha_{_W} < \alpha$, there is a enhancement of 
particles accumulation at the wall, and in the limit of perfectly sticky boundaries ($\alpha_{_W}=0$)
all the particles are asymptotically stuck at walls, i.e. $W_{\infty}=1/2$
(half of particles at each boundary point).
Repulsive walls are instead obtained considering $\alpha_{_W}> \alpha$, resulting in a
depletion effect around the boundaries. In the ideal case of totally repellent walls
($\alpha_{_W} \to \infty$) particles are not able to accumulate at boundary points and $W_{\infty}=0$.
The corresponding expressions for the wall pressure $P$ are easy obtained from the above results
by using the relation $P=NW v/\mu$.


\section{The escape problem}

In the present section we analyze 
escape problems in the presence of confining boundaries.
The case of run-and-tumble particles in a box with partially
reflecting boundaries has been analyzed in a previous paper \cite{Ang_2015}.
Here we extend the investigation to the more realistic situation 
in which particles can get stuck at semi-permeable walls.
The analysis is similar to that of the previous sections, but now we consider 
that walls are partially permeable, allowing particles to escape from the confining region.
Boundary conditions are again given by Eq.s (\ref{eq_br},\ref{eq_bl}), 
but the continuity equations for $W$ - Eq.s (\ref{eq_wb},\ref{eq_wa}) -   now read
\begin{eqnarray}
\label{eq_ewb}
\partial_t W_b(t) &=& J(b,t) - \lambda W_b(t)\\  
\label{eq_ewa}
\partial_t W_a(t) &=& -J(a,t)- \lambda W_a(t)
\end{eqnarray}
where $\lambda$ is the escape rate, i.e. the rate at which particles stuck at the boundaries escape out of the box.
The boundary conditions for ${\tilde \pr}$, Eq.s (\ref{eq_bpr},\ref{eq_bpl}), are then modified as follow
\begin{eqnarray}
\label{eq_bpr_e}
v  \partial_x {\tilde \pr}|_b &=& - g(s) {\tilde \pr}|_b \\  
\label{eq_bpl_e}
v  \partial_x {\tilde \pr}|_a &=&  g(s) {\tilde \pr}|_a  
\end{eqnarray}
where 
\begin{equation}
g(s)=\frac{(s+\alpha)(s+\lambda)}{s+\alpha+\lambda}
\label{g}
\end{equation}
We note that for $\lambda=0$ one has  $g(s)=s$, and the problem reduces to the previous case of impermeable boundaries.
In the opposite limit of totally permeable boundaries, $\lambda \to \infty$,
one has $g(s)=s+\alpha$ and the usual first-passage time problem is recovered \cite{Ang_2015}.

\begin{figure}[t!]
\includegraphics[width=1\linewidth] {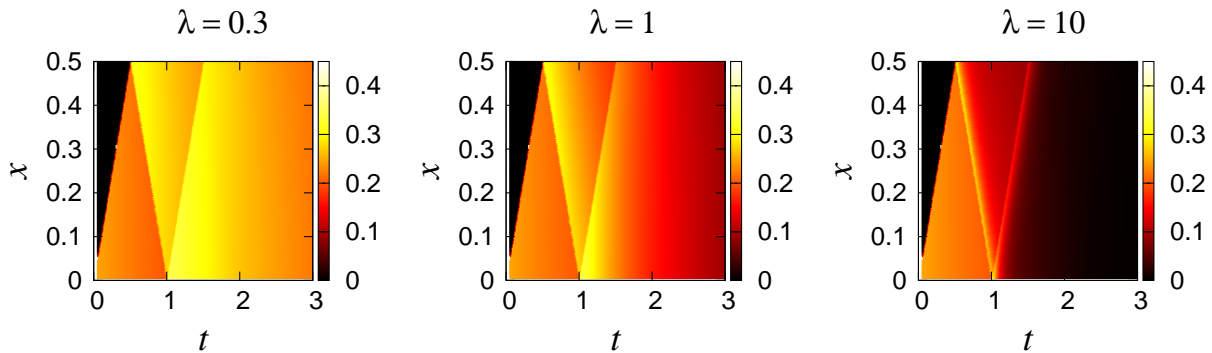}
\caption{\label{fig3}
Probability distribution of particles $\pr$ in the plane 
$(x,t)$ obtained by numerically inverting the Laplace tranformed 
distribution ${\tilde \pr}$  (\ref{Pxs}).
Plots correspond to $L=1$ and (from left to right) $\lambda=0.3, 1, 10$.
Internal units are used: $v=1$, $\alpha=1$, $\mu=1$.
}
\end{figure}

The probability distribution of the escape time $\varphi (t)$ can be obtained from the relation
$\varphi = - \partial_t \mathbb{P}$,
where $\mathbb{P}(t)$ is  the survival probability, i.e. 
the probability that the particle has not yet been absorbed at time $t$,
$\mathbb{P}(t) = \int_a^b dx\ \pr(x,t)+ W_a(t)+W_b(t)$.
In the Laplace domain we have
${\tilde \varphi}(s) = ({\tilde \pr}|_b + {\tilde \pr}|_a) \lambda v /(s+\lambda +\alpha)$
and, by solving Eq. (\ref{eq_lap}) with the new boundary conditions
(\ref{eq_bpr_e})-(\ref{eq_bpl_e}), 
we finally obtain
\begin{equation}
{\tilde \varphi}(s) = \frac{\lambda}{s+\lambda} \  \frac{g}{g \cosh{(cL/2)} + vc \sinh{(cL/2)}}
\label{phis}
\end{equation}
where $c(s)$ and $g(s)$ are given by Eq.s (\ref{c}) and (\ref{g})
and we have considered the symmetric case $b=-a=L/2$.
The fraction of particles at wall is given by ${\tilde W}(s) = 2 \lambda \ {\tilde \varphi}(s)$
and the probability distribution of particles inside the interval is
\begin{equation}
{\tilde \pr}(x,s)=  \frac{c}{2s}\ \frac{ vc \cosh{[c(L/2-|x|)]} + g \sinh{[c(L/2-|x|)]} }
{ g \cosh{(cL/2)} + vc \sinh{(cL/2)} }
\label{Pxs}
\end{equation}
\noindent In Fig. \ref{fig3} the quantity  $\pr (x,t)$, calculated by numerically inverting 
the Laplace tranformed distribution ${\tilde \pr} (x,s)$, is reported for $L=1$ and 
three values of the escape rate, $\lambda=0.3, 1, 10$.
For symmetric reason only the positive half part of the interval is shown, $x>0$. 
Discontinuity lines correspond to the propagation of the initial $\delta$-peaked distribution 
at $x=0$.

\begin{figure}[t!]
\includegraphics[width=0.95\linewidth] {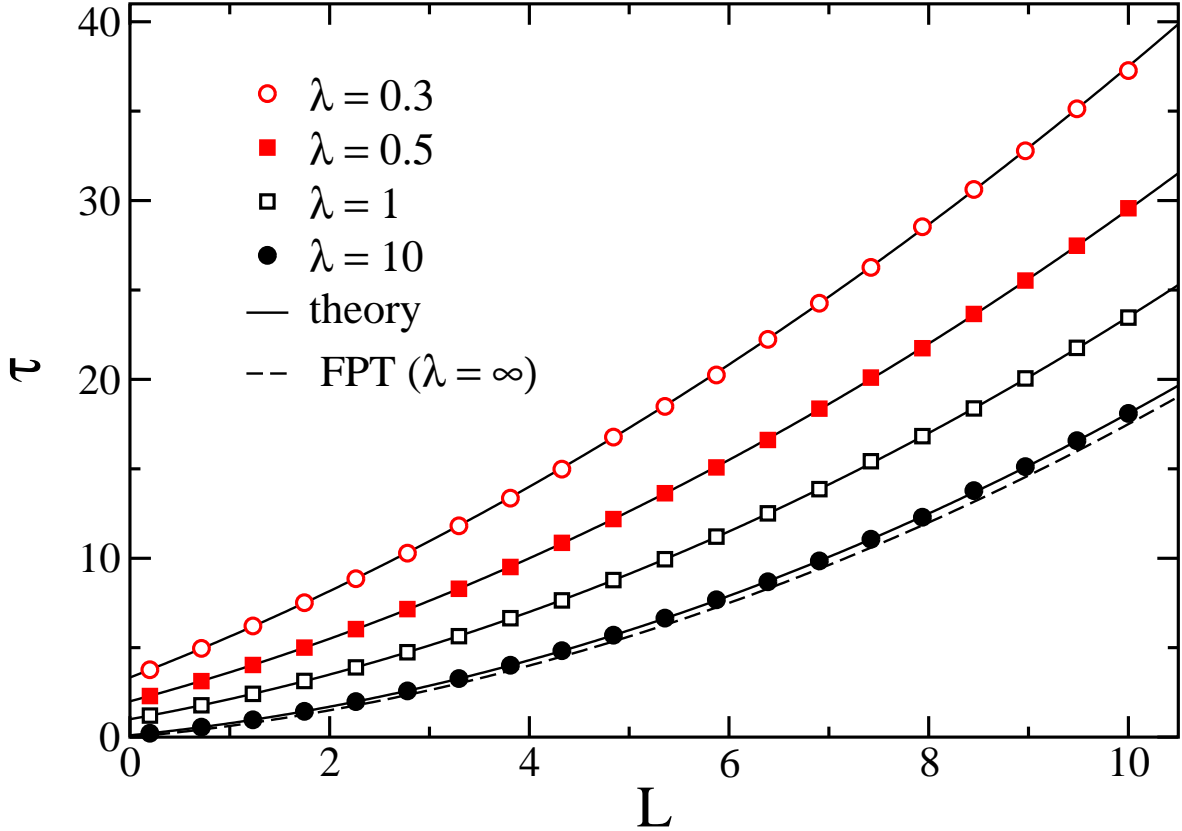}
\caption{\label{fig4}
Escape time $\tau$ as a function of box length $L$ for different escape rates $\lambda$.
Symbols are from simulations and lines from theory, Eq. (\ref{tau}).
Dashed line refers to the mean first-passage-time (FPT), corresponding to $\lambda \to \infty$.
Internal units are used: $v=1$, $\alpha=1$, $\mu=1$.
}
\end{figure}

\noindent The mean escape time is obtained from $\tau = -\partial_s {\tilde \varphi}|_0$ 
\begin{equation}
\tau = \frac{L}{2v} + \frac{\alpha L^2}{8v^2} + \frac{1}{\lambda} \left(1+\frac{\alpha L }{2v}\right)
\label{tau}
\end{equation}
In Fig. \ref{fig4} the mean escape time $\tau$ is reported as a function 
of box length $L$ for different escape rates $\lambda$.
Symbols in the figure are obtained from simulations, performed by using the model 
described in the previous section,
by including that particles stuck at boundaries can escape with probability per unit time $\lambda$.
Regarding swimmers properties one obviously has that lower mean escape times are observed
for fast swimmers with low tumble rates. However it is interesting to note that for
isodiffusive particles, with the same diffusion constant $D=v^2/\alpha$, 
there is an optimal swim speed $v_0=\sqrt{\lambda D}$ which minimizes the mean escape time.

In the case of impermeable boundaries ($\lambda =0$) one has a divergent 
escape time, due to the fact that particles can not escape from the box.
In the opposite limit, considering totally permeable boundaries ($\lambda \to \infty$), 
the expression of the mean first-passage time is recovered, 
$\tau_{_{FPT}} = L/2v + \alpha L^2/8v^2$
\cite{EPJE_2014,Ang_2015}.
We note that in the wave limit (no tumbling, $\alpha=0$) the escape time distribution 
(\ref{phis}) has a simple form
${\tilde \varphi}(s)= (1+s/\lambda)^{-1} \exp(-Ls/2v)$, 
whose inverse-Laplace transform  reads 
$\varphi(t)= \lambda \exp[-\lambda(t-L/2v)] \theta(t-L/2v)$.
In this limit the mean escape time is simply given by the sum of the time needed for a particle to
reach the boundary plus the mean absorption time $\lambda^{-1}$,  $\tau_{no-tumble}=L/2v + 1/\lambda$.\\

\section{Conclusions}

The solution of the telegrapher's equation describing run-and-tumble particles in a finite
1D box is reported.
Suitable boundary conditions are introduced to describe the effects of confining hard walls.
Expressions for  the spatial distribution of particles (Laplace time-transformed)
and  the pressure on the walls are given, focusing on the difference between the initial 
and final pressure on the boundaries exerted by swimmers
starting their motion at the center of the box.
For box length values smaller than few swimmers mean free path the ``wave'' 
character of the swimmers dominates the dynamics and a kind of {\it bounce} 
effect is observed: in their first collision with the wall 
the swimmers exert an higher pressure with respect to the asymptotic one. 
Escape problems are also investigated, which are of interest when active particles are
confined by permeable boundaries.
By introducing a finite probability for swimmers to exit the box, we reported 
the expression for the mean escape time, which reduces to 
the first-passage time in the case of totally permeable walls.

It would be interesting to extend the analysis reported in this work 
to the case of higher spatial dimensions,
considering different kinds of boundary shapes and possible alignment mechanisms.
Moreover, it would be interesting to conceive experiments to validate the 
main results reported in this paper, e.g. the {\it bounce} pressure effect, 
and possibly finding practical implementations, as for example to use the escape time 
as indicator of cell motility for various applications, such as swimmers classification
or sperm selection for in-vitro fertilization.



\end{document}